\documentclass[prc,aastex,amsmath,amssymb,twocolumn,widetext,floatfix,aps,showpacs]{revtex4-2} 

\usepackage{graphicx}
\usepackage{bm}
\usepackage{textcomp}
\usepackage{xcolor}
\definecolor{myblue}{rgb}{0.0, 0.0, 0.6}
\usepackage{hyperref}
\hypersetup{
  colorlinks = true,
  citecolor  = myblue,
  linkcolor  = myblue,
  urlcolor   = myblue
}
\usepackage[utf8]{inputenc}

\begin{document}

\title{
  Breakup corrections to spin asymmetries in the $^3$He beam polarization measurement with the Polarized Atomic Hydrogen Gas Jet Target}%

\author{A.~A.~Poblaguev}\email{poblaguev@bnl.gov}

\affiliation{%
 Brookhaven National Laboratory, Upton, New York 11973, USA
}%
\date{August 2, 2023}

\begin{abstract}
  The requirements for hadron polarimetry at the future Electron-Ion Collider (EIC) include measurements of the absolute helion ($^3$He, $h$) beam polarization with systematic uncertainties better than $\sigma^\text{syst}_P/P\le1\%$. Recently, it was proposed that the Polarized Atomic Hydrogen Gas Jet Target (HJET) be utilized for the precision measurement of the polarization of the $\approx$100\,GeV/n helion beam. At the Relativistic Heavy Ion Collider, HJET serves to determine the absolute proton beam polarization with low systematic uncertainties of about $\delta^\text{syst}P/P\lesssim0.5\%$. To adapt the HJET method for the EIC helion beam, the experimentally determined ratio of the beam and target (jet) spin-correlated asymmetries should be adjusted by the ratio of $p^\uparrow{h}$ and $h^\uparrow{p}$ analyzing powers. A potential problem with the suggested method is that the breakup of $^3$He in polarization measurements could drastically affect the analyzing power ratio. However, an analysis of the breakup corrections, presented in this paper, reveals that while these corrections can be as substantial as $\approx\!4\%$, the effect cancels out to a negligible level in the measured beam polarization.
\end{abstract}

\maketitle

\section{Introduction}

The physics program requirements\,\cite{AbdulKhalek:2021gbh} for hadron polarimetry at the Electron-Ion Collider (EIC)  \cite{Accardi:2012qut} include the precise determination of the $^3$He ($A_h\!=\!3$, $Z_h\!=\!2$) beam polarization,
\begin{equation}
  \sigma_P^\text{syst}/P \lesssim 1\%.
  \label{eq:systEIC}
\end{equation}

It has been advocated\,\cite{Poblaguev:2022gqy,Poblaguev:2022hsi} that the Atomic Polarized Hydrogen Gas Jet Target (HJET)\,\cite{Zelenski:2005mz} can be used for this purpose.

In the Relativistic Heavy Ion Collider, HJET is employed to measure the absolute transverse (vertical) polarization of the proton beams with a low systematic uncertainty of about $\sigma_P^\text{syst}/P \lesssim 0.5\%$\,\cite{Poblaguev:2020qbw}. Recoil protons from the RHIC beam scattering off the jet target are counted in left-right symmetric Si strip detectors. The HJET geometry predetermines the detection of the recoil protons only in the Coulomb-nuclear interference (CNI) scattering constrained by 
\begin{equation}
  0.0013<-t<0.018\,\rm{GeV}^2.
  \label{eq:tRange}
\end{equation}
The Lorentz-invariant momentum transfer $t$ can be simply related to the recoil proton energy $T_R$,
\begin{equation}
  -t = 2m_pT_R,
\end{equation}
where $m_p$ is the recoil particle (proton) mass.

The $^3$He beam polarization $P_h$ can be related\,\cite{Poblaguev:2022gqy} to the precisely known jet target polarization, $P_\text{jet}\!\approx\!0.96\!\pm\!0.001$\,\cite{Zelenski:2005mz}, as
\begin{equation}
  P_\text{meas}(T_R)=P_\text{jet}\frac{a_\text{beam}(T_R)}{a_\text{jet}(T_R)}\times%
  {\cal R}_h(T_R),
  \label{eq:Pmeas}
\end{equation}
where $a_\text{beam}(T_R)$ and $a_\text{jet}(T_R)$ are experimentally determined beam and target (jet) spin asymmetries\,\cite{Poblaguev:2020qbw}, respectively. The ratio of the proton ($p^\uparrow{}h$) and helion ($h^\uparrow{}p$) spin analyzing powers\,\cite{Buttimore:1998rj} is denoted by
\begin{equation}
  {\cal R}_h(T_R) =\frac{A_\text{N}^{ph}(T_R)}{A_\text{N}^{hp}(T_R)} =
  \frac%
  {\kappa_p - 2I_5^{ph} - 2R_5^{ph}\,T_R/T_c}%
  {\kappa_h - 2I_5^{hp} - 2R_5^{hp}\,T_R/T_c},
  \label{eq:Ch}
\end{equation}
where $\kappa_p\!=\!\mu_p\!-\!1\!=\!1.793$ and $\kappa_h\!=\!\mu_h/Z_h\!-\!m_p/m_h\!=\!-1.398$ \cite{Buttimore:2009zz} are derived from the magnetic moments of the proton and helion, $T_c\!=\!4\pi\alpha{Z_h}/m_p\sigma_\text{tot}^{ph}\!\approx\!0.7\,\text{MeV}$, and $\sigma_\text{tot}^{ph}$ is the total $\mathit{ph}$ cross section. The hadronic spin-flip amplitude parameters $r_5\!=\!R_5\!+\!iI_5$ for $p^{\uparrow}h$ and $h^{\uparrow}p$ can be calculated\,\cite{Kopeliovich:2000kz,Buttimore:2001df,Poblaguev:2022hsi} with sufficient accuracy, $r_5^{ph}\!\approx\!r_5^{pp}$ and $r_5^{hp}\!\approx\!r_5^{pp}/3$, using the proton-proton value $r_5^{pp}$ ($|r_5^{pp}|\!\sim\!0.02$) measured at HJET\,\cite{Poblaguev:2019saw}.

For the calculated beam polarization $P_h$, the systematic errors due to possible uncertainties in values of $R_5$ and $T_c$ can be eliminated if the measured polarization $P_\text{meas}(T_R)$ is extrapolated to $T_R\!\to\!0$,
\begin{align}
  P_\text{meas}(T_R) &= P_h\times\left[1+\xi(T_R)\right],  \\
  \xi(T_R) &= \xi_0+\xi_1T_R/T_c+ \ldots.
  \label{eq:xi}
\end{align}

The right-hand side of Eq.\,(\ref{eq:Ch}) displays only the leading-order approximation of the interference terms in the numerator of the CNI elastic analyzing power expression\,\cite{Buttimore:1998rj}:
\begin{equation}
  A_\text{N}(t)\!\;{=}\!\;\frac%
  {-2\operatorname{Im}\left[
    \phi_5^\text{em} \phi_+^\text{had\,*}\!+%
    \phi_5^\text{had}\phi_+^\text{em\,*}\!+%
    \phi_5^\text{had}\phi_+^\text{had\,*}\right]}%
  {|\phi_+^\text{had}\!+\!\phi_+^\text{em}|^2}.
  \label{eq:AN}
\end{equation}
Here, $\phi_5$ and $\phi_+$ are spin-flip and nonflip helicity amplitudes, the hadronic and electromagnetic parts of which are denoted by \lq\lq{had}\rq\rq\ and \lq\lq{em}\rq\rq.
  
$^3$He breakup in the scattering can effectively alter the (elastic) interference terms
\begin{equation}
  \left\{\operatorname{Im}\phi_5\phi_+^*\right\}_I \to%
  \left\{\operatorname{Im}\phi_5\phi_+^*\right\}_I \times\left[1+\omega_I(T_R)\right]
\end{equation}
(schematically discriminated by index $I$) and, consequently, results in a systematic error in the measured $P_h$.

A breakup correction,
\begin{equation}
  \left|\phi_+^\text{had}\right|^2\ \to%
  \left|\phi_+^\text{had}\right|^2\times[1\!+\!\omega(t)],
\end{equation}
to the $pd$ cross section was evaluated in Ref.\,\cite{Poblaguev:2022hsi}, using experimental data obtained in the unpolarized deuteron ($d$) beam measurements in HJET. Extrapolating the result obtained to the $^3$He beam scattering, the breakup related systematic error in measured $P_h$ was found to be negligible.

However, it was underlined in Ref.\,\cite{Poblaguev:2022hsi} that an oversimplified and unjustified theoretical model was used to interpret the deuteron data and to make the extrapolation. Additionally, some assumptions used in the data analysis were not reliably verified.

Although the conclusion drawn in Ref.\,\cite{Poblaguev:2022hsi} was found to be stable against possible variations of the model used, a more comprehensive analysis is necessary for considering HJET in precision ${}3$He polarimetry at EIC.

In this paper, the breakup fraction estimated in Ref.\,\cite{Poblaguev:2022hsi} is compared with the ${dp}$ and ${hp}$ scattering study results in the hydrogen bubble-chamber experiment\,\cite{Aladashvili:1977xe,Dubna-Kosice-Moscow-Strasbourg-Tbilisi-Warsaw:1993lmp}. The HJET and bubble-chamber results were found to be in fair consistency within the experimental accuracy of the measurements. No evidence was found that the breakup fraction given in Ref.\,\cite{Poblaguev:2022hsi} was underestimated therein.

Following basic principles of the Glauber theory\,\cite{Glauber:1955qq,*Glauber:1959}, it can be readily shown that for high-energy polarized proton scattering off a nucleus target, both elastic and breakup, the ratio of hadronic spin-flip and nonflip amplitudes is the same as for $p^{\uparrow}p$ scattering. It also was found that for high-energy forward $p^{\uparrow}h$ and $h^{\uparrow}p$ scattering,  breakup corrections $\omega_I(t)$ to the interference terms $\phi_5^\text{em} \phi_+^\text{had\,*}$, $\phi_5^\text{had}\phi_+^\text{em\,*}$, $\phi_5^\text{had}\phi_+^\text{had\,*}$ are the same, within a relative accuracy of about 10\!\:--\!\:20\%, as the correction $\omega(t)$ to the cross-section term $|\phi_+^\text{had}|^2$. 

The analysis carried out here improves the confidence of the conclusion reached in Ref.\,\cite{Poblaguev:2022hsi} that the EIC $^3$He beam polarization can be measured by HJET with low systematic uncertainty (\ref{eq:systEIC}).

\section{The Breakup Fraction in the HJET ${}^2$H and ${}^3$He Beam Elastic Data}

\subsection{A Model Used to Isolate the Breakup Events}

Considering the breakup $A(p,p)X$ reaction as elastic scattering of a nucleon cluster in the nucleus off the jet proton, one can relate\,\cite{Poblaguev:2022hsi} the missing mass excess $\Delta = m_X - m_A$ to the recoil proton energy:
\begin{equation}
    \Delta = \left(1 - \frac{m^*}{m_A}\right)T_R + p_x\sqrt{\frac{2T_R}{m_p}},
    \label{eq:Delta}
\end{equation}
where $m^*$ is the nucleon cluster mass and $p_x$ is the internal motion's transverse momentum of the cluster in the direction of the detector. Since only low-energy recoil protons ($T_R \lesssim 10\,\text{MeV}$) can be detected at HJET, $\Delta$ is much less than $m_A$. This indicates that the breakup event rate for this process would be strongly suppressed by the phase-space factor.

Obviously, for the deuteron beam, $m^*\!=\!m_p$. Assuming that the $p_x$ distribution in a deuteron is given by a unity integral normalized Breit\:\!--\:\!Wigner function
\begin{equation}
  f_\text{BW}(p_x,\sigma_{p}) = \frac{\pi^{-1}\,\sqrt{2}\sigma_{p}}{p_x^2+2\sigma_{p}^2},
  \label{eq:fBW}
\end{equation}
the $d\to pn$ breakup fraction $\omega(T_R,\Delta)$ can be found\,\cite{Poblaguev:2022hsi} as a convolution of the $\Delta$ distribution, calculated in accordance with Eq.\,(\ref{eq:Delta}), and the phase-space integral (calculated as a function of $\Delta$),  
\begin{align}
  &\omega(T_R,\Delta) =%
  \frac{d^2\sigma_\text{brk}(T_R,\Delta)}{d\sigma_\text{el}(T_R)\,d\Delta} =%
  \frac{\sqrt{2m_pm_n}}{4\pi m_d}
  \nonumber \\ &\qquad\times%
  |\bar{\psi}(T_R,\Delta)|^2f_\text{BW}(\Delta\!-\!\Delta_0,\sigma_\Delta)%
  \sqrt{ \frac{\Delta\!-\!\Delta^d_\text{thr}}{m_d} },
  \label{eq:omegaTR-D} 
  \\  &
  \Delta_0=(1-m_p/m_d)T_R,\qquad%
  \sigma^2_\Delta=2\sigma_{p}^2T_R/m_p.
  \label{eq:omegaD0}
\end{align}
Since the key dependence of the breakup amplitude on $T_R$ and $\Delta$ is allocated in the function  $f_\text{BW}$,  $\bar{\psi}(T_R,\Delta)$ should be interpreted as a reduced ratio of the breakup to elastic amplitudes. Therefore, for HJET measurements, i.e., for low $T_R$ and $\Delta$, $\bar{\psi}(T_R,\Delta)$ can be substituted by a constant $\bar{\psi}\!\equiv\!\bar{\psi}(0,0)$. For the $d\!\to\!pn$ breakup, the threshold is $\Delta^d_\text{thr}\!=\!m_p\!+\!m_n\!-\!m_d\!=\!2.2\,\text{MeV}$.

Integrating over $\Delta$, one can calculate the breakup fraction $\omega(T_R)$, as a function of the recoil proton energy, in the elastic data.

\subsection{The Deuteron Beam Breakup in HJET}

In RHIC Run\,16\, deuteron\:\!--\:\!gold collisions were studied at several beam energies. Since HJET operated in this unpolarized ion Run, the deuteron beam breakup fraction was experimentally evaluated. The measurements had been done in the recoil proton kinetic-energy range $2.8\!<\!T_R\!<\!4.2\,\text{MeV}$, and in the data fit, it was estimated\,\cite{Poblaguev:2022hsi}:
\begin{equation}
  \sigma_p\approx35\,\text{MeV},\qquad  |\bar{\psi}|\approx5.7.
  \label{eq:omegaCalib}
\end{equation}

The value of $\sigma_p$ leads, within the model used, to the following slope of the diffraction cone in the elastic $\mathit{pd}$ scattering
\begin{equation}
  B^{pd} = (1-m_p/m_d)^2/8\sigma_p^2 + B = 37\,\text{GeV}^{-2}
\end{equation}
where $B\!=\!11\,\text{GeV}^{-2}$\,\cite{Bartenev:1973jz,*Bartenev:1973kk} is the elastic $\mathit{pp}$ slope. The calculation is in reasonable agreement with the experimental values for $B^{pd}$\,\cite{Beznogikh:1973uka,Akimov:1975rm}.

\begin{figure}[t]
  \begin{center}
  \includegraphics[width=0.9\columnwidth]{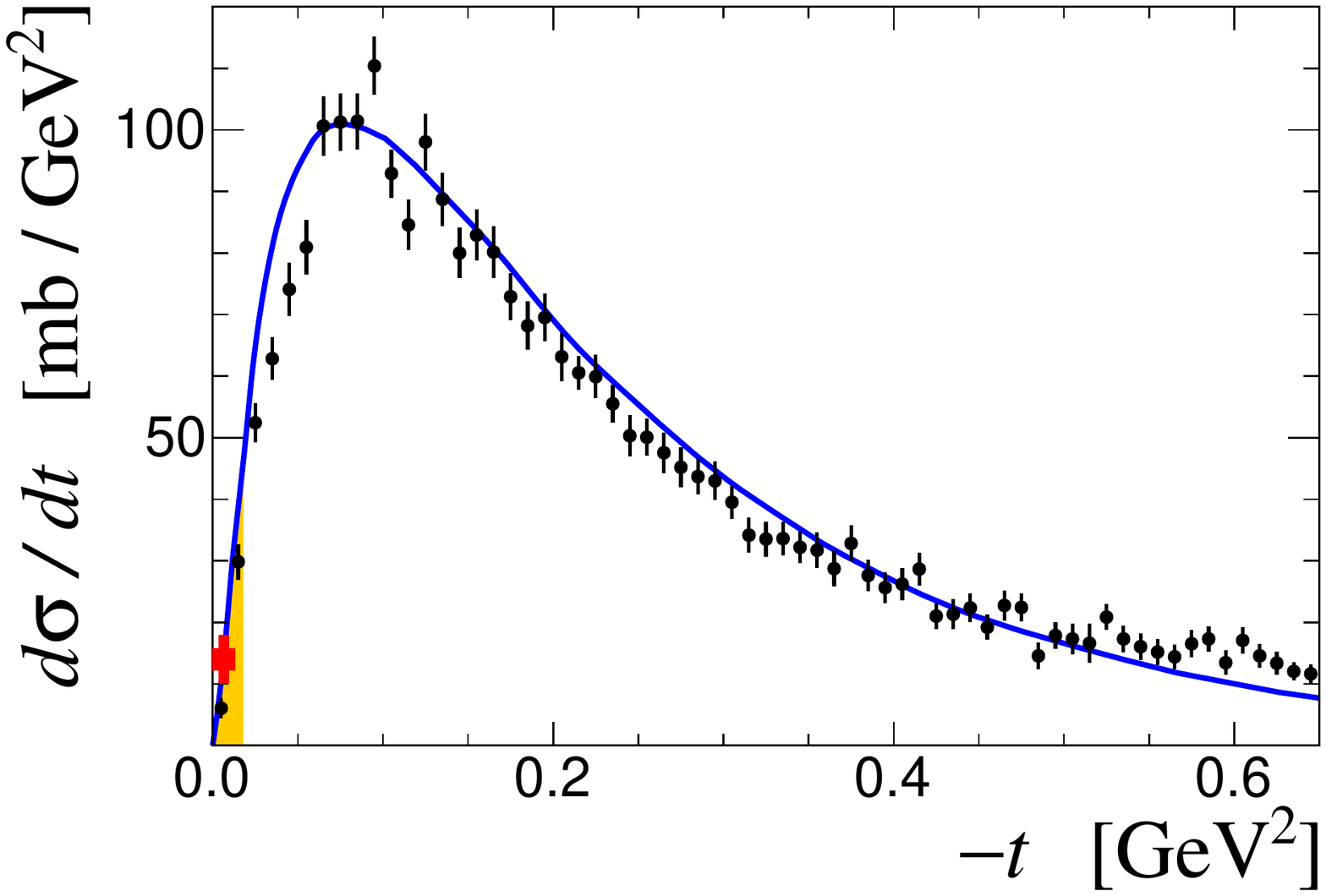}
  \end{center}
  \caption{\label{fig:pd-ppn}
  Differential cross section for the $pd\!\to\!ppn$ breakup scattering. The experimental data ($\bullet$) and theoretical calculation (solid line) are taken from Ref.\,\cite{Aladashvili:1977xe}. The result evaluated in the HJET measurement\,\cite{Poblaguev:2022hsi} is marked {\color{red}$\blacksquare$}. The filled area indicates part of the distribution available for measurement at HJET.
  }
\end{figure}

In the HJET measurements, the breakup fraction in the elastic $\mathit{pd}$ data was estimated\,\cite{Poblaguev:2022hsi} as 
\begin{equation}
  \big\langle\omega_d(T_R)\big\rangle_\text{2.8\:\!--\:\!4.2\,MeV} = 5.0\pm1.4\%.
  \label{eq:dBreakup}
\end{equation}
The value found can be compared with the $pd$\:\!$\to$\:\!$ppn$ differential cross-section (Fig.\,\ref{fig:pd-ppn}) measured in  1.8\,GeV/nucleon deuteron beam scattering in a hydrogen bubble chamber\,\cite{Aladashvili:1977xe}.  Using, for normalization,  elastic $pd$ differential cross section\,\cite{Beznogikh:1973uka}, Eq.\,(\ref{eq:dBreakup}) can be rewritten as
\begin{equation}
  d\sigma/dt\big|_{-t=0.0066\,\text{GeV}^2} = 14\pm4\,\text{mb/GeV}^2.
\end{equation}
The result is in good agreement with the final-state interaction model calculation\,\cite{Aladashvili:1977xe}, $\approx$\:\!$15\,\text{mb/GeV}^2$, and in fair consistency with the value, $8\pm2\,\text{mb/GeV}^2$, interpolated from experimental points in Fig.\,\ref{fig:pd-ppn}. It should also be pointed out that only a small fraction, $\approx$\:\!$1.5\%$, of all  $pd\to ppn$ events can be detected in HJET. 

\subsection{The Helion Beam Breakup in HJET}

Applying the breakup model and using parametrization (\ref{eq:omegaCalib}) for the $^3$He beam, one can evaluate\,\cite{Poblaguev:2022hsi} the breakup fraction for the HJET momentum transfer range 
\begin{equation}
  \big\langle\omega(T_R)\big\rangle_\text{1\:\!--\:\!10\,MeV} = 2.4\pm0.4\,\%, \label{eq:hBreakup}
\end{equation}
while disregarding the detector acceptance and assuming that, for low $t$, only two-body breakup $h\!\to\!pd$ is taken into account, as three-body breakup $h\!\to\!ppn$ is strongly suppressed by a phase-space factor. This result was obtained by substituting $m_d\!\to\!m_h$, $m_n\!\to\!m_d$, and $\Delta_\text{thr}^d\!\to\!\Delta_\text{thr}^h\!=\!5.5\,\text{MeV}$ in Eqs. (\ref{eq:omegaTR-D}) and (\ref{eq:omegaD0}).

\begin{figure}[t]
  \begin{center}
  \includegraphics[width=0.9\columnwidth]{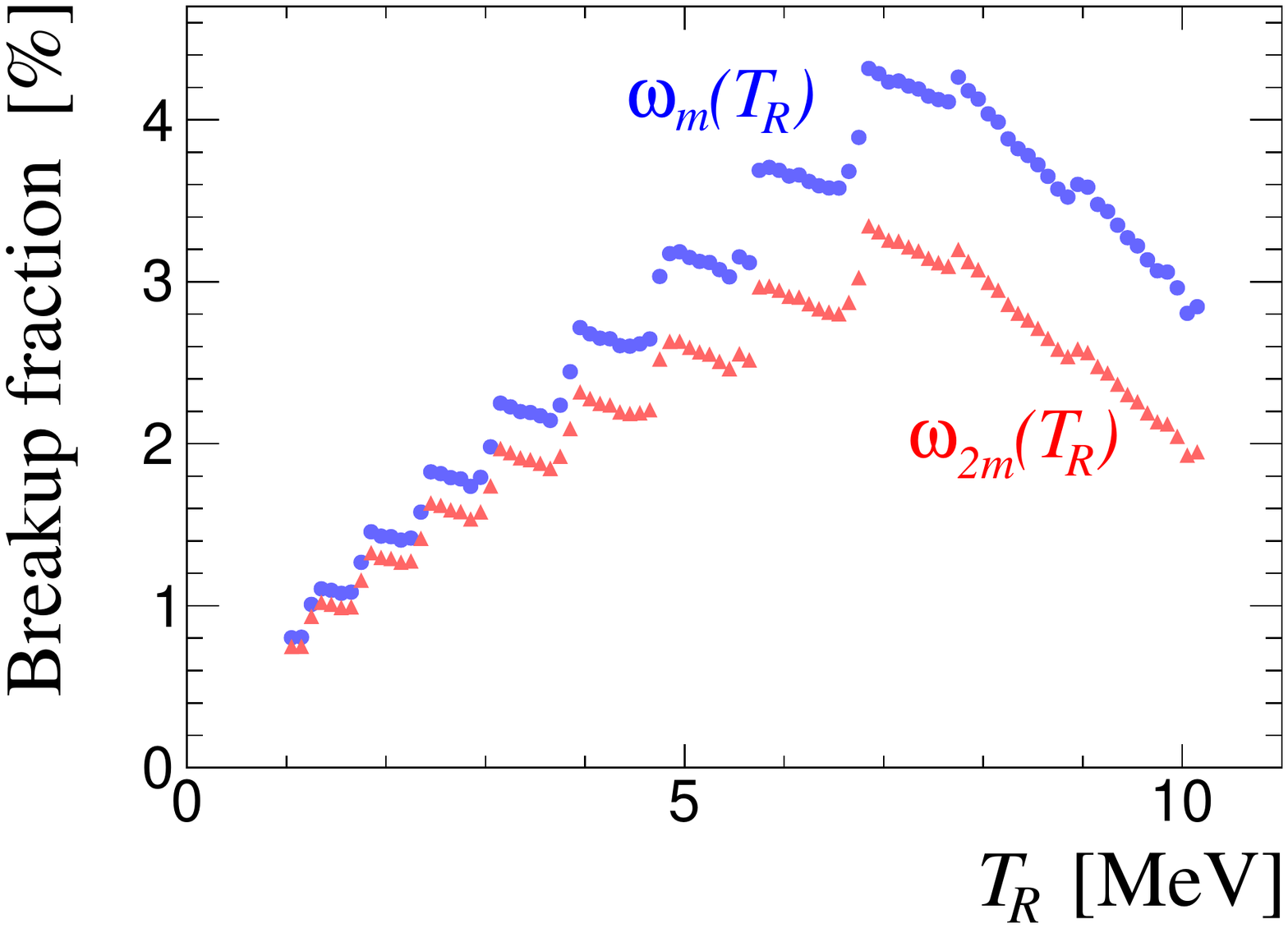}
  \end{center}
  \caption{\label{fig:omega}
    An estimate of the breakup fractions $\omega(T_R)$, corresponding to $m^*\!=\!m_p$ and $m^*\!=\!2m_p$, for 100\,GeV/nucleon $^3$He beam polarization  measurement with HJET. For calculations, the function $f_\text{BW}$ parametrization (\ref{eq:omegaCalib}) found in the deuteron beam data analysis\,\cite{Poblaguev:2022hsi} was used.
  }
\end{figure}

Breakup fraction $\omega_m(T_R)$, calculated for the 100\,GeV/nucleon helion beam and $m^*\!=\!m_p$, is depicted in Fig.\,\ref{fig:omega}. The nonsmooth dependence of the calculated points on $T_R$ reflects the discrete changes in the event selection efficiency attributed to the Si strip width. Near 7\,MeV, the linear dependence on $T_R$ is broken due to the finite size of the detector.

\begin{figure}[t]
  \begin{center}
  \includegraphics[width=0.9\columnwidth]{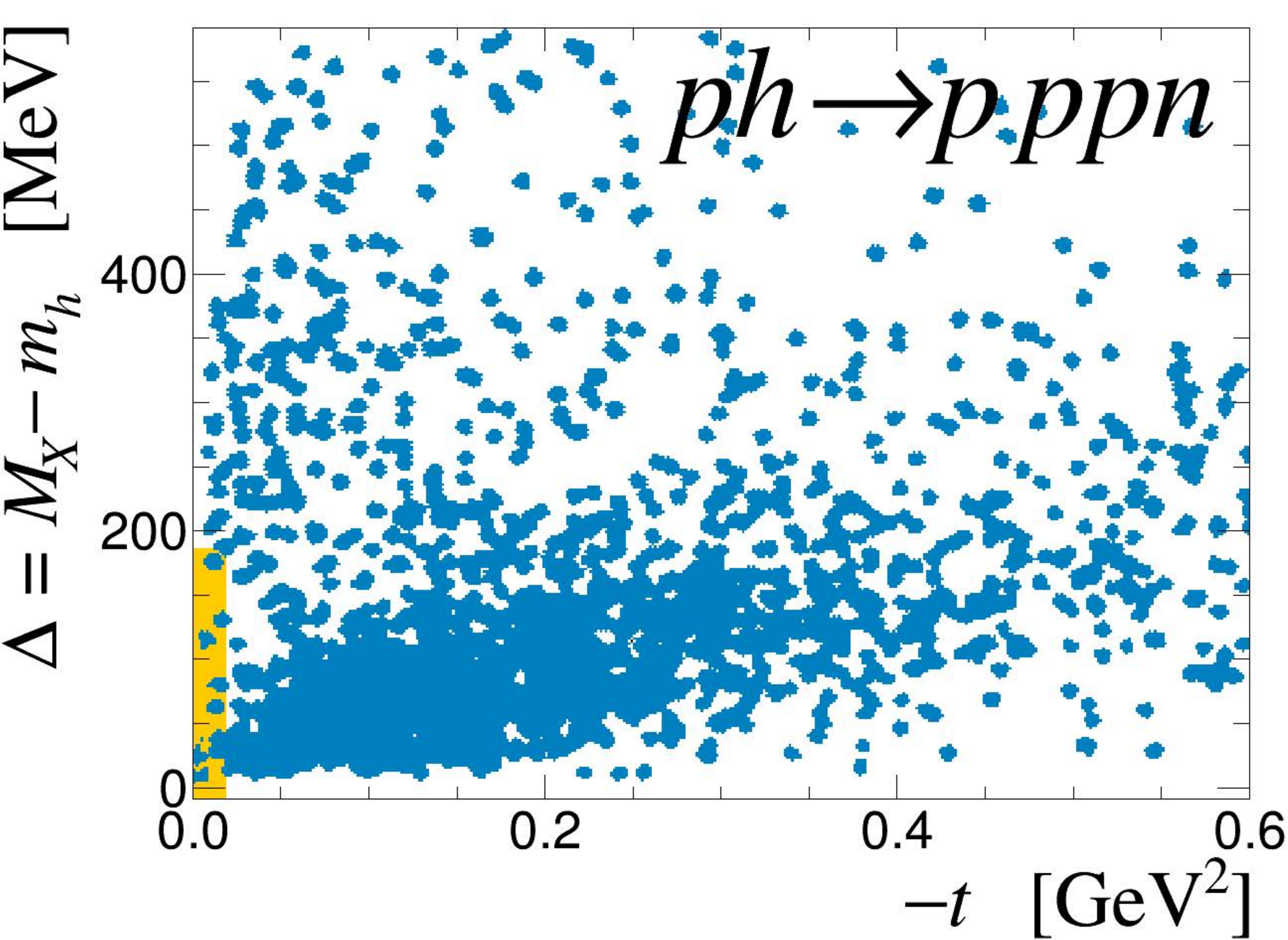}
  \includegraphics[width=0.9\columnwidth]{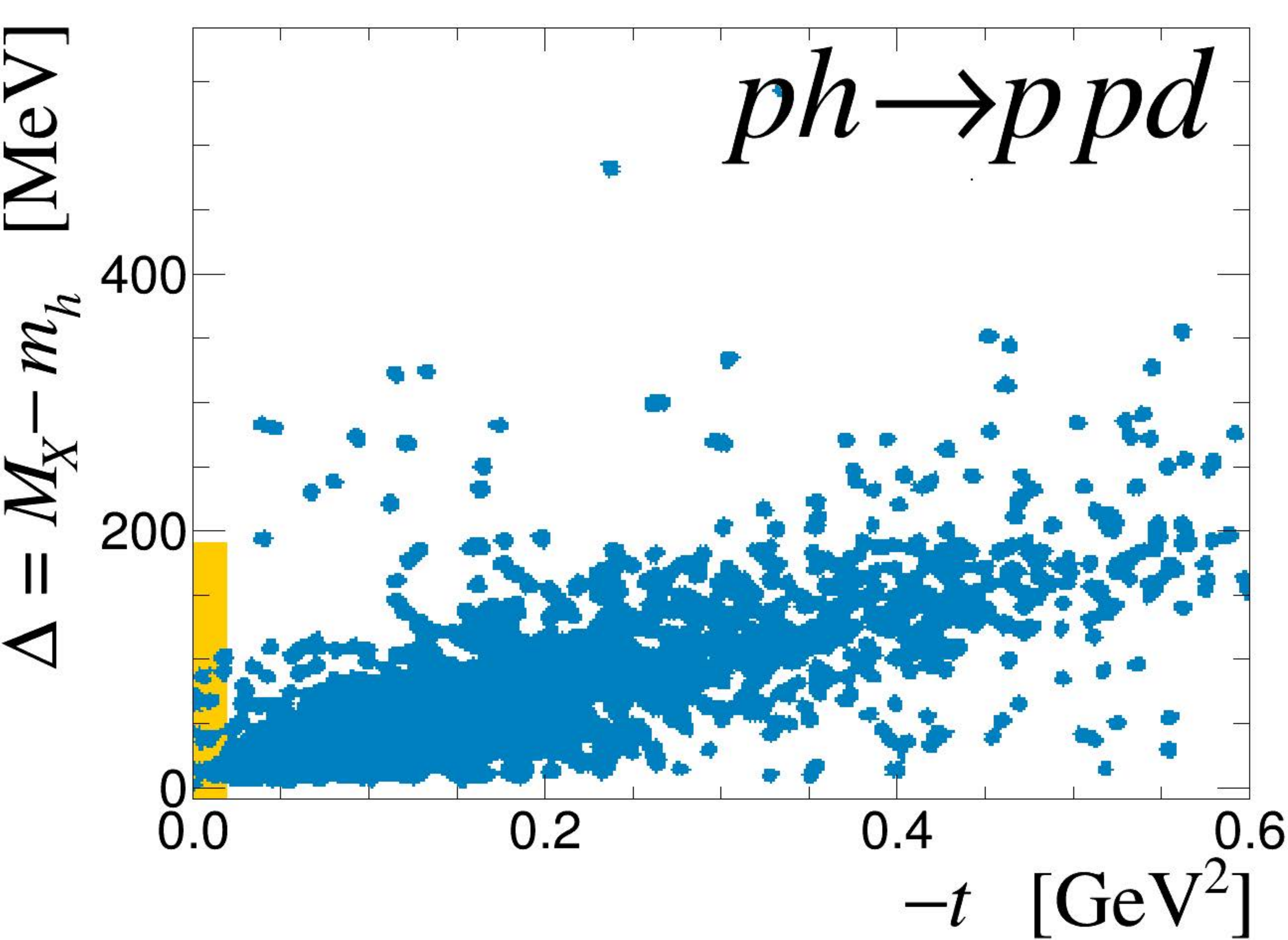}
  \end{center}
  \caption{\label{fig:ph}
    Scattered mass excess $\Delta$ versus momentum transfer plots for helion 4.6\,GeV/nucleon charge-retention breakup scattering, $ph\!\to\!pppn$ and $ph\!\to\!ppd$, measured in the hydrogen bubble-chamber experiment\,\cite{Stepaniak:1996sn}.  The filled areas indicate part of the distribution that can be studied at HJET. }
\end{figure}

  Since, for forward $h\!\to\!pd$ breakup scattering, the $^3$He approximation by a bound state of proton and deuteron cannot be excluded a priori, the breakup fraction $\omega_{2m}(T_R)$ (corresponding to a coherent scattering off two nucleons) should also be considered. Furthermore, the $m^*\!=\!2m_p$ approximation may be relevant, due to the long-distance nature of the Coulomb interaction, for the $\phi_+^\text{em}$ breakup amplitudes.
  
In further analysis, it will be assumed that all of the $^3$He functions $\omega(T_R)$ and $\omega_I(T_R)$ are constrained by $\omega_{m}(T_R)$ and $\omega_{2m}(T_R)$. To tolerate a possible discrepancy between $^3$He breakup corrections, the following function
\begin{equation}
  \Delta\omega(T_R) = \omega_{m}(T_R)-\omega_{2m}(T_R)
\end{equation}
will be used.

For a 4.6\,GeV/nucleon $hp$ scattering, the elastic, $\sigma_{h\to h}\!=\!24.2\!\pm\!1.0\,\text{mb}$, and breakup, $\sigma_{h\to pd}\!=\!7.29\!\pm\!0.14$\,mb and $\sigma_{h\to ppn}\!=\!6.90\!\pm\!0.14$\,mb, cross sections were determined in the hydrogen bubble-chamber measurements\,\cite{Dubna-Kosice-Moscow-Strasbourg-Tbilisi-Warsaw:1993lmp}. For the HJET momentum transfer range (\ref{eq:tRange}), the effective elastic cross section can be derived from the measured $\sigma_{h\to h}$:
\begin{equation}
  \sigma_{h\to h}^\text{HJET} \approx11\,\text{mb}.
\end{equation}

The scattered mass $M_X$ versus momentum transfer $t$ plots (see Fig.\,\ref{fig:ph}) for the breakup scattering were presented in Ref.\,\cite{Stepaniak:1996sn}. It may be pointed out that the correlation seen is in a {\em qualitative} agreement with Eq.\,(\ref{eq:Delta}). It was underlined\,\cite{Stepaniak:1996sn} that the breakup events band is spread around the line
\begin{equation}
  M_X^2 = m_h^2 -2t,
\end{equation}
which is the same as that followed from Eq.\,(\ref{eq:Delta}) if $m^*\!=\!m_p$ and $p_x\!=\!0$.

Each event in Fig.\,\ref{fig:ph} plots contributes nearly 0.003\,mb to the corresponding cross section. For $h\!\to\!ppn$, the breakup band events can be counted in the HJET momentum transfer range. After applying corrections due to the recoil proton detection efficiency at low $t$\,\cite{DUBNA-WARSAW:1975mxc}, one arrives at
\begin{equation}
  \sigma_{h\to ppn}^\text{HJET} < 0.02\,\text{mb}.
\end{equation}
Since it is not unreasonable that mainly background events were counted, the result obtained should be interpreted as an upper limit.

Assuming that $h\!\to\!pd$ breakup has a flat $d\sigma/dt$ distribution in the $0<-t<0.45\,\text{GeV}^2$ momentum transfer range, one immediately finds $\sigma_{h\to pd}^\text{HJET}/\sigma_{h\to pd}\approx0.04$. However, since $\omega(t)\to0$ if $t\to0$, a correction is needed. Guessing that the correction is the same as for the $h\to pd$ distribution in Fig.\,\ref{fig:pd-ppn}, the effective cross-section can be estimated as
\begin{equation}
  \sigma_{h\to pd}^\text{HJET} \sim 0.15\,\text{mb}.
\end{equation}
For comparison, the result displayed in  Eq.\,(\ref{eq:hBreakup}) corresponds to 0.22\!\:--\!\:0.25\,mb depending of value of $m^*$ used.

Thus, the following conclusion comes after Ref.\,\cite{Stepaniak:1996sn}:\\
\noindent{--~~} the helion beam breakup events which can be detected at HJET are mostly $h\to pd$;\\
\noindent{--~~} $\omega_m(T_R)$ depicted in Fig.\,\ref{fig:omega} should be interpreted as an upper limit for the helion beam breakup in HJET.

\begin{figure}[t]
  \begin{center}
  \includegraphics[width=0.73\columnwidth]{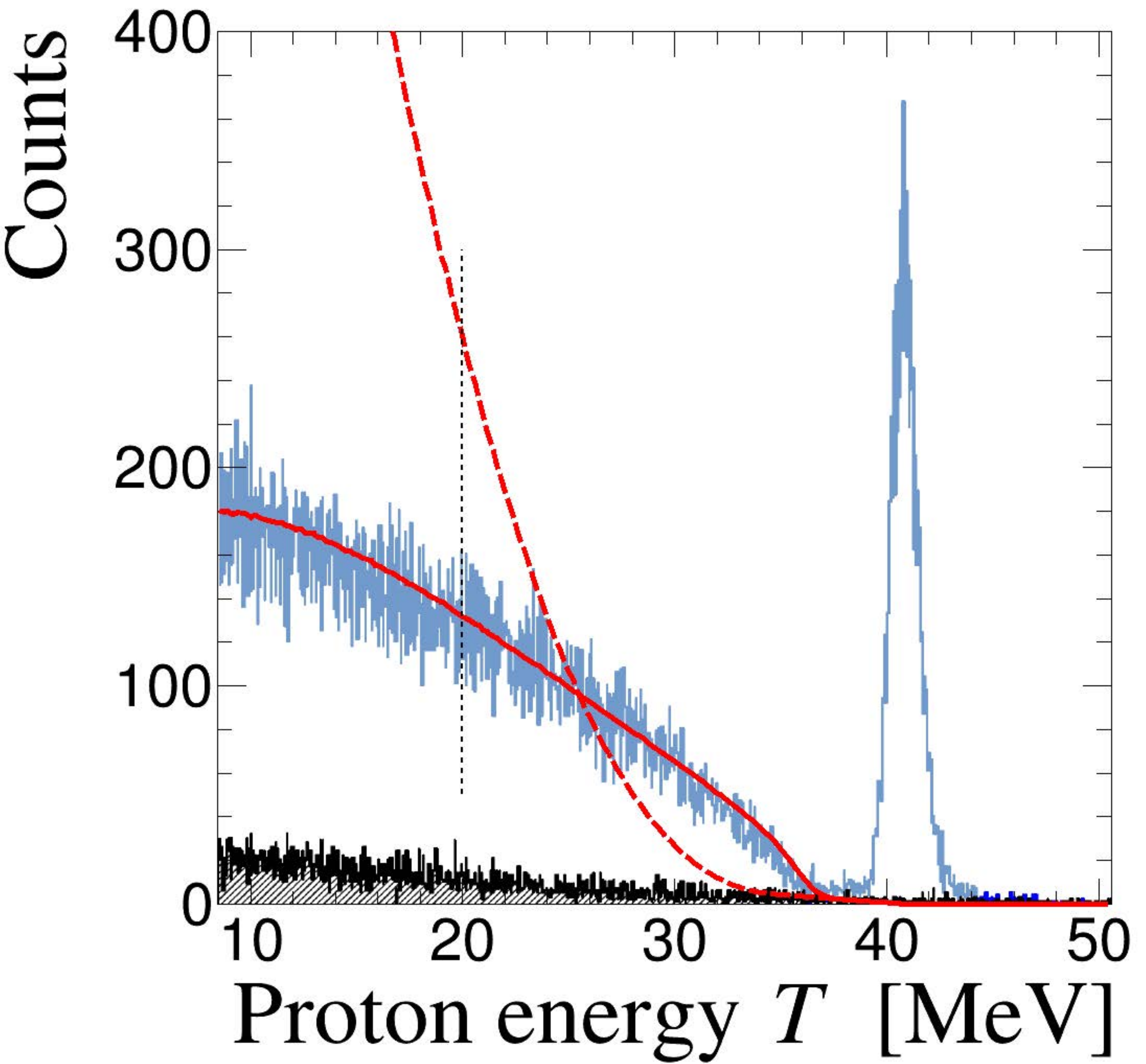}
  \end{center}
  \caption{\label{fig:65MeV}
    Light output spectrum of scattered protons obtained by
    the NaI(Tl) scintillator at $\theta_\text{lab}\!=\!70^\circ$ in the 65\,MeV $p\,{}^3\text{He}$ elastic-scattering study\,\cite{Watanabe:2021jee}. The hatched region indicates the background contribution. Estimates (done here) for the $h\!\to\!{pd}$ and $h\!\to\!{ppn}$ breakup components are shown by solid and dashed red lines, respectively. The calculations were normalized by the total number of the breakup protons with energy above 20\,MeV ($\Delta\!<\!24\,\text{MeV}$).
  }
\end{figure}

Another evidence of the $h\!\to\!ppn$ breakup suppression for low $\Delta$ follows from the scattered proton energy spectrum (Fig.\,\ref{fig:65MeV}) measured in the 65\,MeV proton scattering off the $^3$He target\,\cite{Watanabe:2021jee}. The elastic events are centered at $T_\text{el}\!=\!40.8\,\text{MeV}$. Since energy measured in the breakup scattering can be roughly approached by $T_\text{brk}\!\approx\!T_\text{el}\!-\!\Delta$,  an evident square root dependence on $T$ at the breakup endpoint should be attributed to the phase space integral,  $\propto$\:\!$(\Delta\!-\!\Delta_\text{thr}^h)^{1/2}$, in the two-body breakup event rate in Eq.\,(\ref{eq:omegaTR-D}).

Applying Eqs.\,(\ref{eq:omegaTR-D})\,and\,(\ref{eq:omegaD0}) to the $h\!\to\!pd$ breakup and using $\sigma_p\!=\!90\,\text{MeV}$, one can fairly well approximate the experimental spectrum in Fig.\,\ref{fig:65MeV}. Since low energy, 65\,MeV, scattering is considered, the effective values of $\sigma_p$ and $|\bar{\psi}|$ may differ from those (\ref{eq:omegaCalib}) evaluated in the high energy, 10\:\!--\:\!31\,GeV, $pd$ scattering.

For the three-body breakup, $h\!\to\!ppn$, phase space integral dependence in the event rate (\ref{eq:omegaTR-D}) is proportional to  $(\Delta\!-\!\Delta_\text{thr}^h)^2$ which makes the calculated spectrum inconsistent with the experimental one. So, there is no evidence of the $h\!\to\!ppn$ events in Fig.\,\ref{fig:65MeV}.

\section{Hadronic Spin-Flip Amplitudes in High-Energy Proton-Nucleus Scattering}

  In Ref.\,\cite{Kopeliovich:2000kz}, it was shown that, at high energy, to a very good approximation, the ratio of the spin-flip to the nonflip parts of the elastic proton-nucleus amplitude is the same as for proton-nucleon scattering. In terms of the hadronic spin-flip amplitude parameter $r_5$, the result can be written as
\begin{equation}
  r_5^{pA} = \frac{i+\rho^{pA}}{i+\rho^{pp}}r_5\approx r_5,
  \label{eq:r5_pA}
\end{equation}
where $\rho^{pA}$ and $\rho^{pp}$ are the real-to-imaginary ratio for the elastic $pA$ and $\mathit{pp}$ scattering, respectively. This result can be readily derived considering the polarized proton scattering off an unpolarized nucleus in Glauber (diffraction) approximation\,\cite{Glauber:1955qq,*Glauber:1959}.

In this approach, the elastic ($f\!=\!i$) and/or breakup ($f\!\ne\!i$) proton-nucleus amplitude can be presented\,\cite{Glauber:1970jm} using the following integral over the impact vector $\bm{b}$:
\begin{align}
  F_{fi}(\bm{q}) &= \frac{ik}{2\pi}\int{d^2\bm{b}\,e^{i\bm{qb}}}\,%
    \prod_{j=1}^A{d^3\bm{r}_j}%
    \nonumber \\ &\qquad\qquad%
    \Psi_f^*(\{r_j\})\Gamma(\bm{b},\bm{s}_1\ldots\bm{s}_A)\Psi_i(\{r_j\})%
    \label{eq:Ffi}
\end{align}
where $\bm{k}$ is the momentum of the incident proton, $\Psi_i$ and $\Psi_f$ are the nucleus's initial and final state wave functions, and $\Gamma(\bm{b},\bm{s}_1\ldots\bm{s}_A)$ is the profile function. The positions of the $A$ nucleons in the nucleus were defined by the vectors $\bm{r}_j,~j\!=\!1,\ldots,A$, and $\bm{s}_j$ are the projections of these vectors on the plane perpendicular to $\bm{k}$.

For a proton-deuteron small angle scattering, the elastic $pd$ amplitude $F_{ii}$ can be approximated\,\cite{Franco:1965wi} by combining the proton-proton $f_p$ and proton-neutron $f_n$ ones as 
\begin{align}
  F_{ii}(\bm{q}) &=
  S(\bm{q}/2)f_n(\bm{q}) +  
  S(\bm{q}/2)f_p(\bm{q})
  \nonumber \\ &+
  \frac{i}{2\pi k}\int{S(\bm{q}')f_n(\bm{q}/2\!+\!\bm{q}')f_p(\bm{q}/2\!-\!\bm{q}')d^2\bf{q}'},
  \label{eq:Fii_pd}
\end{align}
where $S(\bm{q})$ can be interpreted as a deuteron form factor.

To calculate the $p^\uparrow d$ spin-flip amplitude $F_{ii}^\text{sf}$, one can utilize the proton-nucleon one, $f_N^\text{sf}(\bm{q})$,  which, according to the definition of $r_5$\,\cite{Buttimore:1998rj}, is
\begin{equation}
  f_N^\text{sf}(\bm{q})=\frac{\bm{qn}}{m_p}\,\frac{r_5^{pp}}{i+\rho^{pp}}\,f_N(\bm{q})
  = \bm{qn}\,\hat{r}_5^{pp}f(\bm{q}).
\end{equation}
Since $|r_5^{pp}|\,q/m_p\lesssim0.003$ in the HJET measurements, only one nonflip amplitude $f_N$ in each term of Eq.~\ref{eq:Fii_pd} should be replaced by its spin-flip counterpart $f_N^\text{sf}$. In particular,
\begin{align}
  f_pf_n &\to \left[(\bm{q}/2\!+\!\bm{q}')\bm{n}f_p\,f_n + f_p\,(\bm{q}/2\!-\!\bm{q}')\bm{n}f_n\right]\,\hat{r}_5^{pp}
 \nonumber \\ &=
 \bm{qn}\,\hat{r}_5^{pp}\,f_pf_n.
\end{align}
Because each term on the right side of Eq.\,(\ref{eq:Fii_pd}) acquired a factor $\bm{qn}\,\hat{r}_5$, one can readily reach Eq.\,(\ref{eq:r5_pA}).

Generally, elastic $pA$ amplitude can be displayed as 
\begin{align}
  F_{ii}(\bm{q}) &=
  \sum_i{\{{\cal S}_if_i\}} + 
  \sum_{i,j}{\{{\cal S}_{ij}f_if_j\}}
  \nonumber \\ &+ 
  \sum_{i,j,k}{\{{\cal S}_{ijk}f_if_jf_k\}} + \ldots,
  \label{eq:Fii_pA}
\end{align}
where, e.g., for the 3-amplitude terms,
\begin{align}
  \{{\cal S}_{ijk}f_if_jf_k\} &=
  \int{{\cal S}_{ijk}(\bm{q}_i',\bm{q}_j',\bm{q}_k')\,%
    f_i(\bm{q}_i')f_j(\bm{q}_j')f_j(\bm{q}_k')}%
    \nonumber \\ &\times%
    \delta(\bm{q}-\bm{q}_i'-\bm{q}_j'-\bm{q}_k')d^2\bm{q}_i'd^2\bm{q}_j'd^2\bm{q}_k'. 
\end{align}
And similarly for other terms. Thus, no detailed knowledge of the form factor functions ${\cal S}_{i{\ldots}j}$ is needed to prove Eq.\,(\ref{eq:r5_pA}).

\begin{figure}[t]
  \begin{center}
  \includegraphics[width=0.4\columnwidth]{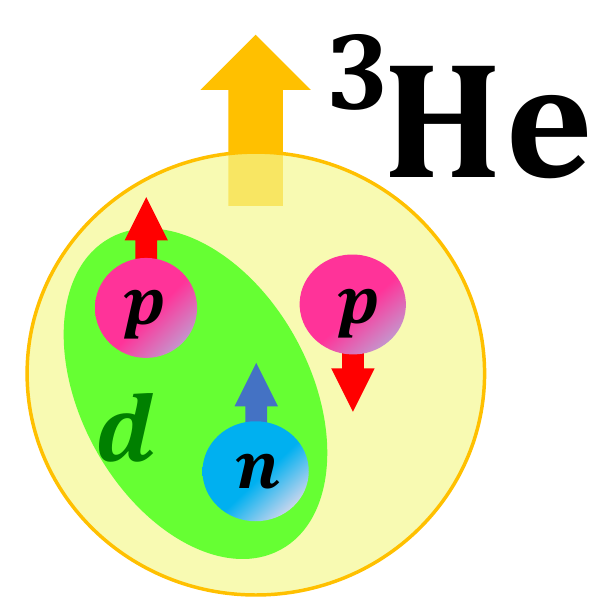}
  \end{center}
  \caption{\label{fig:He3}
    A polarized $^3$He spin structure in the ground ${}^1S_0$ state. Due to the Pauli principle, the protons are in a spin singlet state. A bound state of the neutron and proton with parallel spins can be approximated by a deuteron. 
  }
\end{figure}

In the case of unpolarized proton scattering of a fully polarized nucleus with all nucleons having the same polarization $P$, one can easily find $r_5^{Ap}\!\cong\!r_5^{pp}P$. In the case of a space-symmetric distribution of the nucleons, the result is proportional to the average polarization of the nucleons, $r_5^{pA}\!\cong\!r_5\sum{P_i}/A$. Assuming that $^3$He nuclei are in a space-symmetric ${}^1S_0$ state (Fig.\,\ref{fig:He3}), in which the helion spin is carried by the neutron (i.e., $P_n\!=\!1$ and $P_p\!=\!0$), one can find $r_5^{hp}\!=\!r_5^{pp}/A_h$\,\cite{Buttimore:2001df}. Small corrections to $P_{n,p}$ due to the ${}^3S_1$ and ${}^3D_1$ partial waves were evaluated in Ref.\,\cite{Friar:1990vx}.

Considering a breakup scattering, e.g., $hp\!\to\!dp\:\!p$, one can define the non-flip and spin-flip amplitudes as
\begin{align}
  \!\!F_{fi}(\bm{q}) &= \psi_{fi}(\bm{q})F_{ii}(\bm{q}),\quad
  \psi_{fi}(\bm{q})=|\psi_{fi}(\bm{q})|e^{i\varphi_{fi}(\bm{q})},
  \label{eq:F_fi}
  \\
  \!\!F_{fi}^\text{sf}(\bm{q}) &= \frac{\bm{qn}}{m_p}\,\frac{\widetilde{r}_5}{i+\rho^{pA}}F_{fi}(\bm{q}).
  \label{eq:F_fi_sf}
\end{align}
Here, $F_{fi}(\bm{q})$ and $F_{fi}^\text{sf}(\bm{q})$ are effective breakup amplitudes (for a given value of $\bm{q}$), i.e., sums of all amplitudes over all internal states of the breakup. Therefore, the calculation of probabilities, such as $|F_{fi}|^2$, $|F_{fi}^\text{sf}|^2$, and $\text{Im}\{F_{fi}^\text{sf}F_{fi}^*\}$ assumes summation over these states.

Since, the only difference between elastic and breakup scattering is given [see Eq.\,(\ref{eq:Ffi})] by the final-state wave function $\Psi_f$, the amplitude expansion (\ref{eq:Fii_pA}) should be also valid (with some other set of functions ${\cal S}_{i{\ldots}j}$) for the breakup scattering. Thus,
\begin{equation}
  \widetilde{r}_5^{ph}=r_5^{pp}\frac{i+\rho^{ph}}{i+\rho^{pp}},\quad%
  \widetilde{r}_5^{hp}=(1/3+\delta_{pd})\,r_5^{pp}\frac{i+\rho^{ph}}{i+\rho^{pp}}.
  \label{eq:r5brk}
\end{equation}

Considering a single nucleon scattering of an unpolarized proton from the fully polarized helion $h^\uparrow$ in the ground state (see Fig.\,\ref{fig:He3}) and assuming that $\bm{q}$ is sufficiently large to knock out the target nucleon, one should conclude that the $h\to pd$ helion breakup can occur only if the beam proton was scattered off the oppositely polarized proton, $p^\downarrow$, in $^3$He. Such a speculation suggests that $\widetilde{r}_5^{hp}=-r_5^{pp}$ (or $1/3\!+\!\delta_{pd}\!=\!-1$).

Even for the relatively large value of  $|\delta_{pd}|\!=\!4/3$, the breakup correction to the measured $P_h$ is small\,\cite{Poblaguev:2303.00677}. Moreover, since the accordance between the $h\to pd$ breakup and the beam scattering off the $p^\downarrow$ nucleon should be diluted at low $\bm{q}$, the effect must be much smaller in the HJET measurements. Following discussion in the next section, one can find $\delta_{pd}\!=\!0$ (if $t\!\to\!0$) and
\begin{equation}
  \left|\delta_{pd}/r_5^{hp}\right| \lesssim {\cal O}(1\%)
\end{equation}
in the momentum transfer range (\ref{eq:tRange}).

\section{Evaluation of Breakup Corrections in Helion Beam Polarization Measurements}

A basic assumption of the Glauber theory approach that a high energy beam proton crosses a target nucleus before any changes in the nucleus structure caused by the proton occur, suggests that an electromagnetic interaction is the same for the elastic and breakup scattering. However, to compare the electromagnetic amplitudes for the elastic and breakup scattering, the interaction, in accordance with Eq.\,(\ref{eq:Ffi}), should be projected to the final-state wave functions $\Psi_h$ and $\Psi_{pd}$, respectively. Assuming that the helion disturbance, $^3\text{He}\to{^3\text{He}}^*(\bm{q})$, in the scatterings is small, one finds that the reduced breakup to the elastic ratio (\ref{eq:omegaTR-D}) is the same for the electromagnetic and hadronic as well as for the spin-flip and nonflip amplitudes. 
\begin{equation}
  \frac{\langle\bar{\Psi}_{pd}|{^3\text{He}}^*(\bm{q})\rangle_\text{em}}
       {\langle{\Psi}_{h} |{^3\text{He}}^*(\bm{q})\rangle_\text{em}}
       \approx
  \frac{\langle\bar{\Psi}_{pd}|{^3\text{He}}\rangle_\text{em}}
       {\langle{\Psi}_{h} |{^3\text{He}}\rangle_\text{em}}
       =
  \frac{\langle\bar{\Psi}_{pd}|{^3\text{He}}\rangle_\text{had}}
       {\langle{\Psi}_{h} |{^3\text{He}}\rangle_\text{had}}
       =
       \bar{\psi}.
       \label{eq:ratio}
\end{equation}

For example, since function $f_\text{BW}(\Delta\!-\!\Delta_0,\sigma_\Delta)$ [Eq.\,(\ref{eq:omegaTR-D})] must be the same for the $p^{\uparrow}h$ hadronic spin-flip  and nonflip amplitudes, Eq.\,(\ref{eq:ratio}) immediately leads to $\widetilde{r}_5^{ph}\!=\!r_5^{ph}$ in agreement with Eqs.\,(\ref{eq:r5_pA}) and (\ref{eq:r5brk}).

Would $f_\text{BW}(\Delta\!-\!\Delta_0,\sigma_\Delta)$ be the same for all amplitudes considered in ratio (\ref{eq:Ch}), the breakup correction factors $1\!+\!\omega_I(T_R)$ will cancel in each interference ($I$) term in ${\cal R}_h(T_R)$. In other words, the breakup corrections cannot affect the $^3$He beam polarization measurements in such an approach.

In the more general case, the breakup corrections can be modified (diluted) using the median breakup fraction $\omega^*(T_R)\!=\!\left[\omega_m(T_R)+\omega_{2m}(T_R)\right]/2$:
\begin{equation}
  1+\omega_I(T_R) \to \frac{1+\omega_I(T_R)}{ 1+\omega^*(T_R)}
  = 1+\omega_I'(T_R),
\end{equation}
Since $\omega_I'(T_R)$ is small, of the order of 1\%, 
\begin{equation}
  \left|\omega_I'(T_R)\right|\le\Delta\omega(T_R)/2.
  \label{eq:omega'}
\end{equation}
Replacing the real and imaginary parts of $r_5^{pp}$ in Eq.\,(\ref{eq:Ch}) by the absolute value $|r_5^{pp}|\!\approx\!0.02$ and using the above inequality (\ref{eq:omega'}), one can find {\em conservative} estimates for the breakup corrections $\xi(T_R)$ to the measured polarization separately for each interference term: 
\begin{align}
  \phi_5^\text{em}\phi_+^\text{had}\!:\quad%
  &\xi^\kappa(T_R)%
  <\Delta\omega(T_R) \label{eq:systkappa}
  \\ &\qquad\quad\approx -0.11\%+0.13\%\,\frac{T_R}{T_c},
  \\
  \phi_5^\text{had}\phi_+^\text{em}\!:\quad%
  &\xi^{I_5}(T_R)%
  <\left(\frac{1}{\kappa_p}\!-\!\frac{1}{3\kappa_h}\right)%
  |r_5^{pp}|\,\Delta\omega(T_R)
  \\ &\qquad\quad\approx -0.002\%+0.002\%\,\frac{T_R}{T_c},
  \\
  \phi_5^\text{had}\phi_+^\text{had}\!:\quad%
  &\xi^{R_5}(T_R)%
  <\left(\frac{1}{\kappa_p}\!-\!\frac{1}{3\kappa_h}\right)%
  |r_5^{pp}|\,\Delta\omega(T_R)\frac{T_R}{T_c} \label{eq:systR5}
  \\ &\qquad\quad\approx -0.08\%+0.03\%\,\frac{T_R}{T_c}. \label{eq:systR5fit}
\end{align}
For numerical estimates, a linear fit was done in the recoil proton energy range $2\!<\!T_R\!<\!10\,\text{MeV}$\,\cite{Poblaguev:2020Og}. 

Relatively loose constraints, $\left|\xi_0^{R_5}\right|\!=\!0.08\%$, on possible systematic errors in Eqs.\,(\ref{eq:systR5}), (\ref{eq:systR5fit}) can be explained by the essential nonlinearity of the $\Delta\omega(T_R)T_R/T_c$ function. To improve this, a parabolic function
\begin{equation}
  \xi(T_R) \approx \xi_0 + \xi_1\:\!T_R/T_c + \xi_2(T_R/T_c)^2
  \label{eq:brkPol2}
\end{equation}
can be considered to fit (with $\xi_0\!\equiv\!0$) $P_\text{beam}(T_R)$ in the  $2\!<\!T_R\!<\!7\,\text{GeV}$ energy range. In this case, $\left|\xi_0^{R_5}\right|\!\approx\!0.02\%$. 

\section{Summary}

  In this paper, possible $^3$He breakup related corrections to the EIC $^3$He beam polarization measurements with HJET were reviewed.
  
The breakup fraction in the HJET measurements, evaluated\,\cite{Poblaguev:2022hsi} using the deuteron data, was found to be consistent with a value obtained in the hydrogen bubble-chamber experiment\,\cite{Aladashvili:1977xe}. This result proves that, within the experimental accuracy of the measurements\,\cite{Poblaguev:2022hsi}, {\em(i)} the model used to describe the breakup event rate, $dN/dT_Rd\Delta$, adequately simulates the experimental data and {\em(ii)} the breakup event fraction can be reliably monitored in the HJET measurements using 5\:\!--\:\!20\,GeV $^3$He beams.

In Ref.\,\cite{Poblaguev:2022hsi}, the breakup fraction in the future $^3$He beam polarization measurement at EIC was estimated by extrapolation of the $d\!\to\!pn$ results to the two-body $h\!\to\!pd$ breakup and neglecting the three-body $h\!\to\!ppn$ one. Analyzing experimental distributions displayed in Refs.\,\cite{Stepaniak:1996sn,Watanabe:2021jee}, it was confirmed that the $h\!\to\!ppn$ rate is very small in the HJET measurements.

The evaluated\,\cite{Poblaguev:2022hsi} $^3$He breakup rate  was compared with the results of the $^3$He beam scattering in the hydrogen bubble chamber\,\cite{Dubna-Kosice-Moscow-Strasbourg-Tbilisi-Warsaw:1993lmp,Stepaniak:1996sn}. It was found that the functions $\omega(T_R)$ displayed in Fig.\,\ref{fig:omega} should be interpreted as an upper limit for the breakup fraction in the EIC $^3$He beam scattering in HJET.

It was shown that, within the applicability of the Glauber theory, the ratio of the high-energy proton-nucleus $p^{\uparrow}A$ breakup spin-flip and nonflip amplitudes is the same as for elastic proton-proton scattering. Both for $p^{\uparrow}h$ and $h^{\uparrow}p$ scattering in the HJET momentum transfer range, the ratio of the spin-flip to nonflip amplitude is the same, with a percent accuracy, for the elastic and breakup amplitudes.
  
It was also found that the breakup corrections to the interference terms $\omega_I(T_R)$ and the cross section $\omega(T_R)$ are about the same within 10\!\:--\!\:20\% accuracy, which significantly limits possible alteration of the analyzing power ratio (\ref{eq:Ch}). Although the corrections, $\omega_I(T_R)$, may be as large as 4\% (relative), they do not exceed 1\% in the analyzing power ratio ${\cal R}_h(T_R)$ and are reduced to a negligible level if $P_\text{meas}(T_R)$ is extrapolated to $T_R\!\to\!0$.

Considering the breakup correction cancellation in Eq.\,(\ref{eq:AN}), it should be concluded that the breakup corrections alter the effective analyzing powers, $A_\text{N}^{ph}(t)$ and $A_\text{N}^{hp}$, in the HJET $^3$He beam measurement by no more than 1\%. This estimate is also expected to be applicable for the $p^{\uparrow}A$ analyzing power study\,\cite{Poblaguev:2022xoa} at HJET.

Estimates given in Eqs.\,(\ref{eq:systkappa})\:\!--\:\!(\ref{eq:systR5fit}) verify that the breakup corrections are negligible in the $^3$He beam polarization measurements at HJET\,\cite{Poblaguev:2022hsi}.

\acknowledgements{

  The author would like to thank B. Z. Kopeliovich for useful discussions and acknowledges support from the Office of Nuclear Physics in the Office of Science of the US Department of Energy. This work is authored by an employee of Brookhaven Science Associates, LLC under Contract No.\,DE-SC0012704 with the U.S. Department of Energy.}

%

\end{document}